\documentstyle[epsf]{article}
\makeatletter
\def\setcaption#1{\def\@captype{#1}}
\makeatother

\newcommand{\numu}      {\mbox{$\nu_{\rm \mu}$}}
\newcommand{\nue}       {\mbox{${\nu}_{\rm e}$}}

\newcommand{\Frejus}    {Fr\'{e}jus}
\newcommand{\etal}      {{\it et al.}}
\newcommand{\degree}    {\mbox{$^{\circ}$}}

\begin{document}
{\center \it Revision of 1 Mar. 1998 \\}

\medskip

{\center \Large 
Measurement of a small atmospheric $\nu_\mu/\nu_e$ ratio\\
}

\bigskip

{\center \large The Super-Kamiokande Collaboration\\}


\begin{center}
Y.Fukuda$^a$, T.Hayakawa$^a$, E.Ichihara$^a$, K.Inoue$^a$,
K.Ishihara$^a$, H.Ishino$^a$, Y.Itow$^a$,
T.Kajita$^a$, J.Kameda$^a$, S.Kasuga$^a$
% ...

\footnotesize \it

$^a$Institute for Cosmic Ray Research, University of Tokyo, Tanashi,
Tokyo 188-8502, Japan\\
% ...

\end{center}

    \section*{Abstract}

From an exposure of 25.5~kiloton-years of the Super-Kamiokande detector,
900 muon-like and 983 electron-like single-ring atmospheric neutrino
interactions were detected with momentum $p_e > 100$ MeV/$c$, $p_\mu >
200$ MeV/$c$, and with visible energy less than 1.33 GeV. Using a
detailed Monte Carlo simulation, the ratio
$(\mu/e)_{DATA}/(\mu/e)_{MC}$ was measured to be $0.61 \pm
0.03(stat.) \pm 0.05(sys.)$, consistent with previous results from
the Kamiokande, IMB and Soudan-2 experiments, and smaller than
expected from theoretical models of atmospheric neutrino production.

    \section*{Introduction} 

Atmospheric neutrinos are the decay products of hadronic showers
produced by cosmic ray interactions in the atmosphere.  In recent
years, the ratio $R\equiv(\mu/e)_{DATA}/(\mu/e)_{MC}$ has been
measured to study the atmospheric neutrino flavor ratio
$(\nu_\mu+\overline{\nu}_\mu)/(\nu_e+\overline{\nu}_e)$; the ratio
of data to Monte Carlo is taken to cancel uncertainties in the
neutrino flux and cross sections. Here, $(\mu/e)$ denotes the ratio of
the numbers of $\mu$-like to $e$-like neutrino interactions observed
in the data or predicted by the Monte Carlo (MC). The expected value
for $R$ is unity if there is agreement between the experiment and the
theoretical prediction.  The water Cherenkov detectors
Kamiokande~\cite{ref:kam1} and IMB~\cite{ref:imb} have observed a
statistically significant low value of $R$ for ``sub-GeV'' events with
lepton energies of about 1 GeV or less.  The NUSEX~\cite{ref:nusex}
and \Frejus{}~\cite{ref:fre} experiments have reported no deviation
from unity, but with smaller data samples.  Recently the Soudan-2
experiment~\cite{ref:soudan} has also observed an $R$ value smaller
than unity.  Kamiokande~\cite{ref:kam} also observed a smaller $\mu/e$
ratio in the ``multi-GeV'' energy range, as well as a dependence of this
ratio on the zenith angle, and hence the neutrino travel distance. The
small value of the ratio and the zenith angle dependence suggest that
neutrino oscillations may be responsible for these results.

This letter presents the first measurement of the $\mu/e$ ratio using the
Super-Kamiokande detector. The data were restricted to the sub-GeV
range, which comprises contained events with visible energy
less than 1.33 GeV and electron (muon) momentum greater than 100 MeV/$c$
(200 MeV/$c$); these criteria match the definition used by
Kamiokande.\footnote{This analysis has also been performed with a
minimum $p_\mu$ of $300$ MeV/$c$ and $p < 1.5$ GeV, corresponding to the
kinematic cuts used in prior analyses by the IMB
experiment\cite{ref:imb}; results were very similar.} We carried out
two independent analyses (A and B) whose results were consistent with
each other, confirming the validity of the methods used.

    \section*{Super-Kamiokande detector}
Super-Kamiokande is a 50-kiloton water Cherenkov detector located near
the Kamiokande detector, in the Mozumi mine of the Kamioka Mining
Company in Gifu prefecture, Japan.  It lies at a mean overburden of
2,700 meters-water-equivalent below the peak of Mt. Ikenoyama.  The inner
detector comprises 11,146 Hamamatsu R3600 50-cm diameter
photomultiplier tubes (PMT), viewing a cylindrical volume of pure
water 16.9 m in radius and 36.2 m high.  The 50-cm PMTs were specially
designed \cite{ref:idpmt} to have good single photoelectron (p.e.)
response, with timing resolution 2.5 ns RMS. An outer layer of water
2.6 to 2.75~m thick completely surrounds the inner detector to
passively shield against radioactivity from the surrounding rock. The
two detector regions are optically separated by a pair of opaque
sheets which enclose a dead region 55 cm in thickness. The outer
detector is instrumented with 1,885 outward-facing Hamamatsu R1408
20-cm PMTs. In order to increase light collection, 60 cm $\times$ 60 cm
wavelength shifters~\cite{ref:wls} were attached to the outer PMTs,
and all surfaces of the outer detector were covered with reflective
white DuPont Tyvek material. The outer detector data were used to identify
incoming cosmic rays and exiting muons from neutrino interactions.

Both inner and outer PMT signals were processed by asynchronous,
self-triggering circuits that record the time and charge of each PMT
hit over a threshold. The inner PMT signals were digitized with custom
Analog Timing Modules (ATMs)\cite{ref:atm} which provide 1.2 $\mu$s
range at $0.3$ ns resolution in time and $550$ pC range at $0.2$ pC
resolution ($~0.1$ p.e.)  in charge for each PMT. The ATM has
automatically switched dual channels to provide deadtime-free data
acquisition.  The outer PMT signals were processed with custom
charge-to-time conversion modules, which output timing pulses of width
linearly proportional to the integrated charge of the PMT pulse. These
signals were digitized with LeCroy 1877 multi-hit TDCs using $16$ $\mu$s
full range.

A trigger was formed by the coincidence of at least 30 PMT hits in a
200 ns window, over a threshold of about $1/4$ p.e. per PMT.  This
trigger condition corresponds to the mean number of hit PMTs for a
5.7~MeV electron. The trigger rate was 10-12~Hz. The trigger rate due
to cosmic ray muons was 2.2~Hz. Digitized data were saved at a total
rate of 12~GB per day.

Water transparency was measured using a dye laser and CCD camera, and
found to be about 100~m at wavelength 420~nm.  During the time period
described here (approximately 17 months of detector operation), water
transparency was monitored continuously by cosmic-ray muons; the
average effective attenuation length for Cherenkov light increased by
25\%, due to improvement in water clarity resulting from operation of
the water purification system.

The calibration of digitized PMT data to number of p.e.s and arrival
time was performed by offline processes directly linked to the
detector data stream via local network. Both of the independent
analyses began with the same calibrated data. Each analysis
independently estimated the conversion from p.e.s to visible energy
($E_{vis}$), which is defined as the energy of an electromagnetic shower
which produces an equivalent amount of Cherenkov light.  Approximately
9 p.e.s were measured for one MeV of visible energy.  The accuracy
of the absolute energy scale was estimated to be $\pm2.4\%$ based on
several calibration sources: cosmic ray through-going muons, stopping
muons, muon-decay electrons, the invariant mass of $\pi^0$s produced by
neutrino interactions, radioactive source calibration, and a 5-16~MeV
electron LINAC. The estimated momentum resolution for electrons and
muons is $2.5\%/\sqrt{\rm E(GeV)}+0.5\%$ and $3\%$, respectively.

Both analyses required the vertex of the neutrino interactions to be
reconstructed inside a fiducial volume 2~m from the light barrier just
outside the inner PMT plane. This comprised a concentric cylindrical
volume 32.2~m high and 14.9~m in radius with a mass of 22.5~kilotons. 

    \begin{table}[ht]
 \renewcommand{\arraystretch}{1.5}
 \newcommand{\lw}[1]{\smash{\lower2.ex\hbox{#1}}}
 \begin{center}
  \begin{tabular}{lrcrrrr} \hline\hline
     & \lw{Data}\hfil &&\multicolumn{4}{c}{Monte Carlo} \\
     \cline{4-7}
     & & &total    &\nue\  CC(q.e.) & \numu\  CC(q.e.) & NC\\
     \hline
     single-ring   &  1883 && 2030.5 & 720.1(562.4)& 1185.0(921.4)&  125.3 \\
     {} $e$-like     &   983 &&  812.2 & 714.3(558.4)& 18.6  (  4.5)&  79.3 \\
     {} $\mu$-like &   900 &&  1218.3 & 5.8  (  4.0)& 1166.5(916.9)&  46.0 \\
     multi-ring    &   696 &&  759.2 & 182.1( 46.6)& 325.5( 47.3)& 251.6 \\
     \hline
     total         &  2579 && 2789.7 & 902.2(609.0)&1510.5(968.7)& 376.9 \\
     \hline\hline
  \end{tabular}
 \end{center}
 \caption{Summary of the sub-GeV experimental data compared with
the Monte Carlo estimation. Monte Carlo statistics have been
normalized to the live time of the experimental data. ``q.e.'' refers to
quasi-elastic events.} 
 \label{tb:sum}
\end{table}
\begin{table}[ht]
 \renewcommand{\arraystretch}{1.5}
 \newcommand{\ce}[1]{ \multicolumn{1}{c}{#1}}
 \begin{center}
  \begin{tabular}{crrcrr} \hline\hline
                    & \multicolumn{2}{c}{1 or more muon decays}&& \multicolumn{2}{c}{2 or more muon decays}\\
                      \cline{2-3}                \cline{5-6}
                    & \ce{data}& \ce{MC}       && \ce{data} & \ce{MC} \\
   \hline
$\mu$-like & $608/900=67.6\pm1.6$\%&$68.1\pm0.1\pm1.0$\%&&$26/900=2.9\pm0.6$\%&$4.1\pm0.1\pm0.2$\%\\
$e$-like     & $ 91/983= \makebox[0.5em]{}9.3\pm0.9$\%&$8.7\pm0.3\pm0.1$\%&& $2/983=0.2\pm0.1$\%&$0.1\pm0.1\pm0.01$\%\\
    \hline\hline
  \end{tabular}
 \end{center}
 \caption{Percentages of events with muon decay in single-ring events.
The first error value shown is statistical.
For Monte Carlo, the second error value is from the estimated 
muon decay detection efficiency.}
 \label{tb:fraction}
\end{table}

    \section*{Analysis A}
For analysis A, we used data from a 25.5 kiloton-years net exposure,
collected during the period between May 1996 and October 1997.  The
main backgrounds for the observation of atmospheric neutrino events
were cosmic ray muons and low-energy radioactivity in the
detector. These two backgrounds were rejected by requiring no correlated
hits in the outer detector and a minimum deposited energy of 30 MeV in
the inner detector, respectively.

Starting from $\sim 400$ million triggers, the data sample was reduced
to about 12,000 events by applying the following requirements: (1) no
significant outer detector activity (total number of hits less than
25, and no spatial cluster with more than 10 hits), (2) total charge
collected in the inner detector $> 200$ p.e.s, which corresponds to
22~MeV/$c$ for electrons and 190 MeV/$c$ for muons, (3) the ratio
(maximum p.e. in any single PMT)/(total p.e.s) is less than $0.5$, and
(4) the time interval from the preceding event $> 100~\mu$s, to reject
electrons from stopping muon decays. Additional selection criteria
were used to eliminate spurious events, such as those due to
``flashing'' PMTs that emit light from internal corona discharges. The
selected events were hand-scanned by two independent scanners, to
reject remaining background events.  About 6,000 events were
classified as fully-contained events, a large fraction of which were
neutrino interactions with no charged particles exiting into the outer
detector.

The vertex position of an event was determined using PMT hit times;
the point which yielded the sharpest distribution of PMT times
adjusted for the time of flight of Cherenkov light was defined as the
vertex position. The vertex was reconstructed again after particle
identification to correct for particle track length. The vertex
resolution was estimated to be 30~cm for single-ring fully-contained
events. The number of Cherenkov rings and their directions were
determined automatically by a maximum-likelihood procedure. The
efficiency for identifying quasi-elastic $\nu_e$($\nu_\mu$) events as
single-ring was 93(95)\%. The angular resolution for single-ring
events was estimated to be 3~degrees. The momentum of a particle was
determined from the total number of p.e.s within a 70\degree\ 
half-angle cone relative to the track direction, with corrections for
light attenuation and PMT angular acceptance.

\begin{figure}[ht]
   \begin{center}
     \epsfxsize=15cm
     \epsfbox{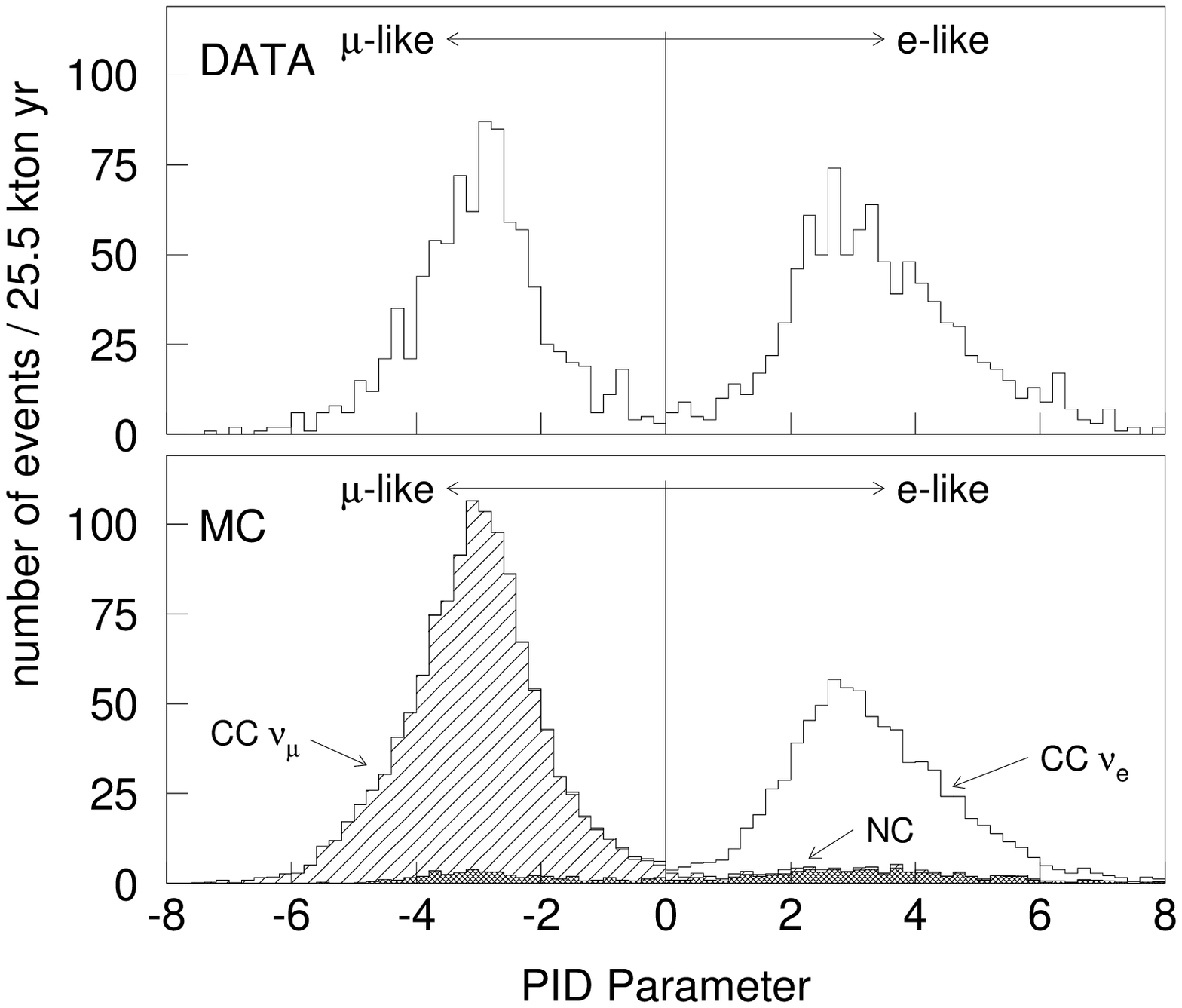}
   \end{center}
   \caption{
   Distribution of the particle identification (PID) parameter
   for single-ring atmospheric neutrino events for
   both data and Monte Carlo samples in Analysis A.
   If the PID parameter of an event is positive (negative),
   the event is classified as $e$-like ($\mu$-like).
   For the Monte Carlo, the contributions from
   charged current and the neutral current events
    are also shown.}
   \label{fig:pidon}
\end{figure}

The particle identification of the final state lepton exploits
systematic differences in the shape and the opening angle of Cherenkov
rings produced by electrons and muons.  Cherenkov rings from
electromagnetic cascades exhibit a more diffuse light distribution
than those from muons. The opening angle of the Cherenkov cone, which
depends on $\beta(\equiv v/c)$, was used to separate electrons and
muons at low momenta.  The validity of the method was confirmed by a
beam test experiment at KEK\cite{ref:beam}.  Figure~\ref{fig:pidon}
shows distributions of the PID parameter (effectively a log-likelihood
difference for the electron and muon hypotheses) for the data and for
Monte Carlo single-ring events. If the PID parameter was
positive(negative), the event was classified as $e$-like($\mu$-like).
The misidentification probabilities for single-ring muons and
electrons were estimated to be $0.5\pm0.1$\% and $0.7\pm0.1$\%,
respectively, using simulated charged-current (CC) quasi-elastic
neutrino events.  The identification efficiency was checked using
cosmic-ray muons which stop in the detector and subsequently decay to
electrons. The resulting misidentification probabilities for these
muon and electron events were $0.4\pm0.1$\% and $1.8\pm0.5$\%,
respectively, in good agreement with the Monte Carlo estimates.  This
check was performed continuously during data-taking, and particle
identification performance remained stable despite increasing water
transparency.

There are several calculations of the expected atmospheric neutrino
flux at the Super-Kamiokande site.  The calculated flux of
Ref.\cite{ref:honda} was used for the Monte Carlo simulation of
atmospheric neutrino interactions.  The neutrino interaction model
took into account quasi-elastic scattering\cite{ref:Llew72-qe},
single-pion production\cite{ref:reinseghal}, coherent pion
production\cite{ref:reinseghal2}, and multi-pion
production\cite{ref:MC}. Propagation of produced leptons and hadrons
was modeled using a GEANT\cite{ref:GEANT}-based detector simulation,
which included Cherenkov light production and propagation in water.
Hadronic interactions were simulated by CALOR\cite{ref:CALOR}, except
for pions with momentum less than 500~MeV/$c$, for which a special
program\cite{ref:MC} was developed, with cross-sections taken from the
experimental results. For pions produced in $^{16}$O nuclei, inelastic
interactions, charge exchange, and absorption in the nuclei were also
taken into account\cite{ref:MC}.  A sample equivalent to 10 years of
detector operation was generated with the Monte Carlo simulator.  This
Monte Carlo sample was then passed through the same event
filtering\footnote{The Monte Carlo events were not hand scanned, except for
selected samples for studies.} and reconstruction as the experimental
data.

From the initial 6,000 fully-contained events, 3,462 neutrino event
candidates were reconstructed in the fiducial volume with $E_{vis}>30$
MeV. We estimated that 83.0\% of the total charged current interaction
events in the fiducial volume were retained in the present sample. The
sources of inefficiency were: non-fully-contained(9.3\%), $E_{vis}$
lower than 30~MeV(5.8\%), reduction inefficiency(0.1\%), and a small
systematic bias toward fitting the vertex position outside of the
fiducial volume (2.1\% and 1.5\% for the $e$-like and $\mu$-like
events respectively).

To measure the $\mu$/$e$ ratio we required that there be only a
single ring identified in the event.  The sub-GeV kinematic
requirements were: $E_{vis}$ less than 1.33~GeV, and electron and muon
momenta greater than 100 and 200~MeV/$c$, respectively.
Table~\ref{tb:sum} summarizes the number of observed events and
compares them with the Monte Carlo estimation.  From these data, we
obtained:
\begin{eqnarray*}
      R  \equiv (\mu/e)_{DATA}/(\mu/e)_{MC} =
      0.61 \pm 0.03(stat.) \pm 0.05(sys.).
\end{eqnarray*}

Sources of systematic uncertainty in $R$ were estimated as follows:
5\% from uncertainty in the predicted \numu/\nue\ flux ratio,
3.5\% from uncertainty in the CC neutrino interaction cross sections
and nuclear effects in the H$_{2}$O target,
3\% from the neutral current (NC) cross section, 
0.5\% from the uncertainty in pion propagation in water,
3\% from single-ring event selection,
2\% from particle misidentification,
1\% from the absolute energy calibration, 
0.6\% from the vertex fit and fiducial volume cut,
less than 0.5\% from contamination by cosmic ray muons, flashing PMT events
and neutron interactions in the detector,
and 1.5\% from statistical uncertainty in the Monte Carlo.
Adding these errors in quadrature, the total systematic
uncertainty is 8\%.

The result using particle identification was checked with the
rate of muon decays in the neutrino events.  The detection efficiency
for muon decay was estimated to be 80\% for $\mu^+$ and 63\% for
$\mu^-$ by a Monte Carlo study.  These figures were confirmed with an
accuracy of 1.5\% using cosmic-ray stopping muons. The fraction of
events with muon decays in the single-ring event sample is shown in
Table \ref{tb:fraction}, and is in good agreement with the Monte Carlo
estimation, for both $\mu$-like and $e$-like events. This supports the
reliability of the particle identification and the Monte Carlo
estimation of pion production.


\newcommand{\livetimeB}{419.0}

\newcommand{\exposureB}{25.8}

\newcommand{\mcexposureB}{10.2} 

\newcommand{\RB}{0.65}  

\newcommand{\RBstat}{0.03}  

\newcommand{\RBsys}{0.05}   

\newcommand{\eventsB}{3,521}

\newcommand{\subGeVB}{2650}

\newcommand{\singlB}{2008}

\newcommand{\mlikeB}{1041}

\newcommand{\elikeB}{967}

\newcommand{\multiB}{642} 

\newcommand{\qeB}{78\%}
\newcommand{\muqeB}{98\%}
\newcommand{\eqeB}{93\%}

\newcommand{\muplusdkeff}{70\%}
\newcommand{\muminusdkeff}{57\%}

\newcommand{\ewithdkB}{$8.4 \pm 0.9\%$} 
\newcommand{\mcewithdkB}{$10.6 \pm 0.4 (stat.) \pm 0.3 (sys.)\%$}

\newcommand{\mwithdkB}{$55.2 \pm 1.5\%$}
\newcommand{\mcmwithdkB}{$55.7 \pm 0.5 (stat.) \pm 1.7 (sys.)\%$}

\newcommand{\mwithdksB}{$2.3 \pm 0.5\%$}
\newcommand{\mcmwithdksB}{$3.3 \pm 0.2 (stat.) \pm 0.1 (sys.)\%$}

\newcommand{\effB}{85\%}   
\newcommand{\meffB}{80\%}  
\newcommand{\eeffB}{95\%}  
\newcommand{\efidB}{2.7\%} 
\newcommand{\mfidB}{2.0\%}

\newcommand{\esys}{1\%}         
\newcommand{\psys}{1.5\%}       
\newcommand{\energysys}{1.8\%}  
\newcommand{\fidsys}{1.5\%}     
\newcommand{\ringsys}{1.9\%}    
\newcommand{\pidsys}{3.5\%}     
\newcommand{\contsys}{0.5\%}    
\newcommand{\detectorsys}{4.6\%} 

\newcommand{\muesys}{5\%}       
\newcommand{\qesys}{3\%}        
\newcommand{\nqesys}{1.6\%}     
\newcommand{\ccsys}{3.4\%}      
\newcommand{\ncsys}{2.2\%}
\newcommand{\mcstat}{1.5\%}     
\newcommand{\mcsys}{6.6\%}      

\newcommand{\totalsys}{8.1\%}   

\section*{Analysis B}

An independent analysis of the Super-Kamiokande data was performed, to
detect possible errors and provide a comparison of reduction and
reconstruction techniques. The computer programs used were completely
independent from Analysis A, as were the determinations of energy scale
and systematic uncertainty. The common starting point for each
analysis was the raw data with electronics calibrations applied.  The
selection of the data sample was slightly different, and analysis B
had an exposure of \exposureB{} kiloton-years.

In Analysis B, the initial set of events was obtained by applying the
following requirements:
(1) fewer than 10 PMT hits in the outer detector 
in a 200 ns window around the trigger time,
(2) total charge collected in the inner detector $>100$ p.e.s,
within a 200 ns time window,
(3) the ratio (maximum p.e.s in any single PMT) / (total
p.e.s) was less than 0.4,
(4) the time interval from the preceding event $>100~\mu$s.

In the next stage of the analysis, a vertex point fit was done by
$\chi^2$ minimization of the difference between the PMT time and the
time expected, based on light propagating from a vertex.  The charge
in a $\pm20$~ns window of residual time was required to be greater
than 150~p.e.s. A second fit was applied for the hypothesis of an
entering muon; if the entry point had more than 2 outer detector tube
hits within 20~m and $\pm$50~ns, the event was rejected.

Remaining flashing PMT events were removed by imposing requirements on
the shape of the residual time distribution. In addition, a separate
analysis was performed that rejected events with repetitive light
patterns characteristic of specific flashing PMTs.  In the final stage
of the reduction, a precise vertex and direction fit was applied. The
vertex reconstruction had 40 cm resolution superimposed on an
uncorrected shift of +43(-43) cm for electrons (muons).

These requirements resulted in \eventsB{} fully contained events
within the fiducial volume.  Among them, we visually identified 3
events caused by electronics noise, 1 flashing PMT event, and 1 event
likely to be an entering cosmic ray muon; however, no events were
removed based on scanning. This constituted a background of 0.2\%,
which was accounted for in the systematic uncertainty.  Based on Monte
Carlo studies we estimated that \effB{} of the total CC
interaction events were retained in the sample:
\meffB{} from $\nu_\mu$ and \eeffB{} from $\nu_e$.  There was a small
systematic bias toward fitting the vertex position for electrons
outside (\efidB{}) and muons inside (2.0\%) of the fiducial volume.

Single-ring events were selected based on cuts using the azimuthal
distribution of light which falls outside of the Cherenkov cone of the
track; events with azimuthal symmetry were considered single-ring, and
those with asymmetry were considered multi-ring. Based on Monte Carlo
studies, the percentage of quasi-elastic interactions in the sample
selected by this algorithm was estimated to be \qeB{} with a
\eqeB{}(\muqeB{}) efficiency for identifying quasi-elastic
$\nu_e$($\nu_\mu$) events as single-ring.

Particle identification was performed using the vertex and direction
from the final track fit and the distribution of PMT charge projected
onto the track axis at the Cherenkov angle.  The shape of this
distribution was used to determine the particle type, primarily by
measuring more projected charge behind the vertex for electromagnetic
showers than for muons. The particle misidentification probabilities
for quasi-elastic $\nu_\mu$ and $\nu_e$ events were
$1.4^{+1.6}_{-0.5}$\% and $3.5^{+1.4}_{-1.5}$\%, respectively.

\begin{figure}[ht]
\begin{center}
\epsfxsize=15cm
\epsfbox{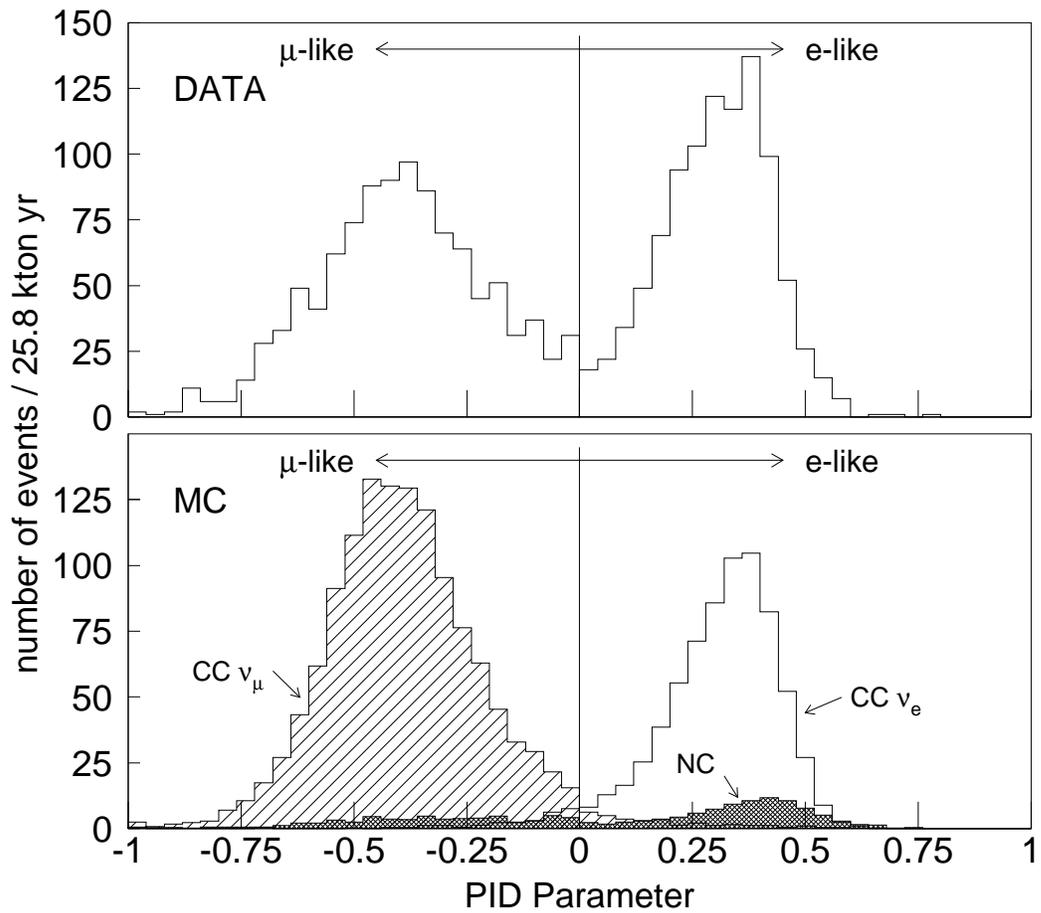}
\caption{\label{fig:adpid} 
   Distribution of the PID parameter used in Analysis B
    for $\mu$-like events (PID$<0$)
   and $e$-like events (PID$>0$) in both data and Monte Carlo
   samples.}
\end{center}
\end{figure}

The momentum of the final state lepton was determined from the total
number of p.e.s in a 20~ns window of residual time, taking into
account the higher Cherenkov threshold for muons. The final data
sample of single-ring events, within the same sub-GeV kinematic range
defined in Analysis A, and with vertices in the fiducial volume,
consisted of 1,041 $\mu$-like events and \elikeB{} $e$-like events.  The
distributions of the particle identification parameter for data and
Monte Carlo are shown in Figure~\ref{fig:adpid}.

A sample of Monte Carlo events was generated corresponding to
\mcexposureB{} years of exposure, using the atmospheric $\nu_\mu$ and
$\nu_e$ flux predictions of Ref.~\cite{ref:gaisser} and the
pion-production models of Rein and
Seghal\cite{ref:reinseghal,ref:reinseghal2} as adapted for use in the
IMB experiment\cite{ref:casper}.  A second sample was generated using
the same flux but following the pion-production model of Fogli and
Nardulli\cite{ref:foglinardulli,ref:hainesatmmc} as a check;
essentially all results were found to agree with the first Monte Carlo
sample within the estimates of systematic uncertainty. Both Monte
Carlo samples used common code to track particles in water, generate
Cherenkov light, and simulate the detector response. The Monte Carlo
events were processed through the same analysis chain as the
experimental data. The classification of data and Monte Carlo events
is summarized in Table~\ref{tb:sumoff}.  Finally, we obtained 
$R$~$= \RB{} \pm \RBstat{}(stat.) \pm \RBsys{}(sys.)$, in good
agreement with analysis A.

Estimated contributions to the systematic uncertainty in $R$ were as
follows: 
\muesys{} from the uncertainty in the predicted $\nu_\mu/\nu_e$ flux ratio,
\pidsys{} from particle misidentification, 
\ccsys{} from the uncertainty in the CC neutrino cross section, 
\ncsys{} from the NC neutrino cross sections, 
\ringsys{} from single-ring selection, 
\energysys{} from the energy calibration,
\fidsys{} from fiducial volume determination, 
less than \contsys{} from non-neutrino backgrounds contamination,
and \mcstat{} from the statistical error in the Monte Carlo. These sum in
quadrature to a total systematic uncertainty of
\totalsys{}.


As in analysis A, the fraction of events with one or more
muon decays was in good agreement with Monte Carlo prediction.
For $e$-like events, \ewithdkB{} were found with one or more decay
signals compared to the Monte Carlo prediction of \mcewithdkB{}. For
$\mu$-like events, \mwithdkB{} were found with one or more decays and
\mwithdksB{} with two or more decays, compared with the Monte Carlo
predictions of \mcmwithdkB{} and \mcmwithdksB{}, respectively.

\begin{table}[ht]
 \renewcommand{\arraystretch}{1.5}
 \newcommand{\lw}[1]{\smash{\lower2.ex\hbox{#1}}}
 \begin{center}
  \begin{tabular}{lrcrrrr} \hline\hline
     & \lw{Data}\hfil &&\multicolumn{4}{c}{Monte Carlo} \\
     \cline{4-7}
     &                        &&total   &\nue{} CC(q.e.)&\numu{} CC(q.e.)& NC\\
     \hline
     single-ring      & \singlB{} && 2185.9 & 724.1(610.7)&1306.6(1095.7)&155.2 \\
     {} $e$-like      & \elikeB{} &&  821.1 & 696.0(587.7)&  32.8(15.4) &  92.4 \\
     {} $\mu$-like    & \mlikeB{} && 1364.8 &  28.1(23.0) &1273.9(1080.3)& 62.8 \\
     multi-ring       & \multiB{} &&  631.3 & 151.1(44.3) & 246.2(25.2) & 233.9 \\
     \hline
     total            & \subGeVB{}&& 2817.2 & 875.3(655.0)&1552.9(1120.9)&389.1 \\
     \hline\hline
  \end{tabular}
 \end{center}
\caption{Summary of sub-GeV events compared with Monte Carlo
estimation, for \exposureB{} kiloton-years of Super-Kamiokande data
processed by Analysis B.}  
\label{tb:sumoff}
\end{table}

    \section*{Comparison of analyses}

Live time selection was slightly different for Analysis~A and
Analysis~B; 95\% of the live time of each group was analyzed by the
other.  For runs analyzed by both groups, we found that 94\% of the
events in the final sample of Analysis~A (both single and multi-ring)
were also included in the final sample of Analysis~B.  Comparing the
ring-counting algorithm of analysis~A with the single-ring selection
cuts of analysis B, we found that 90\% of Analysis A events had the
same classification in Analysis B.  We compared reconstructed
quantities for single-ring events commonly selected by both A and B.
For these events the mean absolute difference in reconstructed vertex
position was 84 cm parallel, and 18 cm perpendicular
to the track direction.  Reconstructed track directions agreed with a
mean of angular difference of 2.5 degrees.  The average difference in
momentum ($\Delta P/P$) was 0.5\%.  Comparing particle identification,
we found that 97\% of Analysis A events agreed with the Analysis B
classification as $e$-like or $\mu$-like.  These results are
consistent with expectation based on the resolution and efficiencies
of the software developed independently by the two analysis groups.
Finally, we found that the systematic uncertainties estimated by each
analysis were consistent with the differences in event reconstruction.


We note that $\mu/e$ ratios for data and Monte Carlo in Analysis A
were smaller than those in Analysis B by 
15\% and 10\%, 
respectively.
The difference in the Monte Carlo ratios is primarily due to
differences in the vertex fitting 
and single-ring selection.
Using live times which did not completely overlap, Analysis A
and B found R values which were different by 5.8\%; however, if common
runs were used the difference was 4.4\%.  This difference is
consistent with the known differences in particle identification
between the two analyses, the systematic error in $R$ due to
different analysis techniques only, and the Monte Carlo statistical error.

    \section*{Results} 
Results from the two independent analyses agree 
well, not only in $R$ but also in all other points of comparison.
Thus, it would be difficult to explain the observed deviation of $R$
from expectation in terms of unresolved mistakes in experimental data
analysis. Since results from the two independent analyses are
consistent, further discussion refers to results from analysis A.

\begin{figure}
   \begin{center}
     \epsfxsize=15cm
     \epsfbox{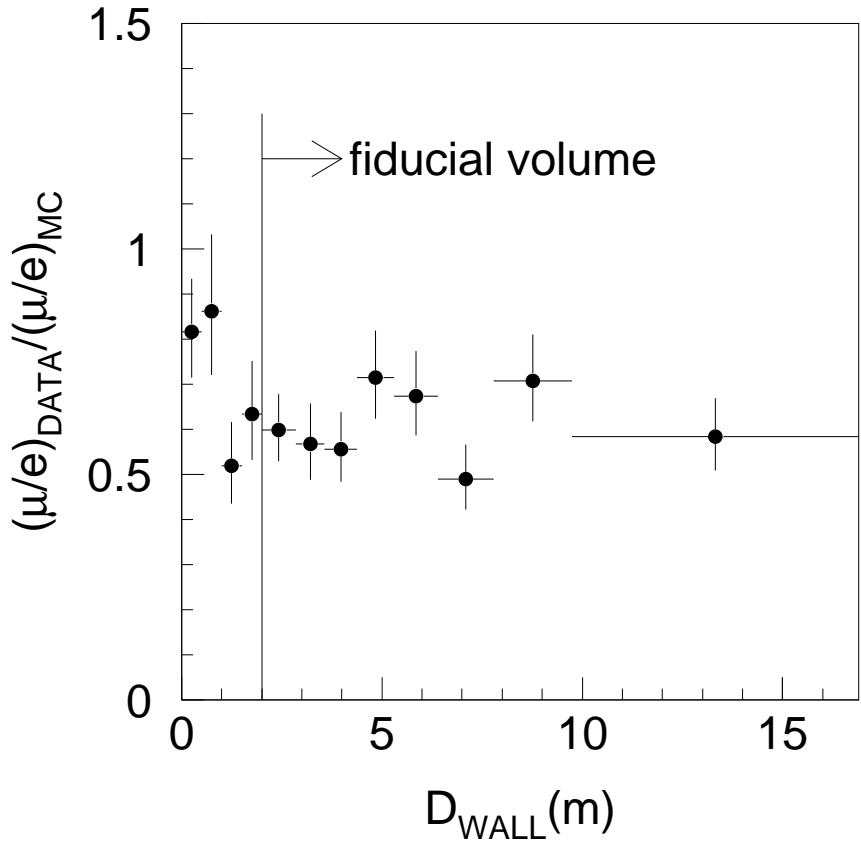}
        
   \end{center}
   \caption{
            $R$ as a function of $D_{WALL}$,
            the distance between the event vertex and inner detector
            wall. The region $D_{WALL}>2$m is the
            fiducial volume.}
   \label{fig:dwall}
\end{figure}

In Figure~\ref{fig:dwall}, $R$ is shown to have no strong dependence
on $D_{WALL}$, the distance from the vertex to the nearest wall (even
outside the fiducial volume at $D_{WALL}<2$ m). There is no evidence
for neutron or other background which could change $(\mu/e)_{DATA}$
near the edge of the fiducial volume. Based on the scanning of events
near $D_{WALL}=0$, we determined that the higher R value in the first
bin was likely to be due to cosmic-ray muon background, but no
significant muon background was observed for the other bins.

\begin{figure}[htb]
   \begin{center}
     \epsfxsize=10cm
     \epsfbox{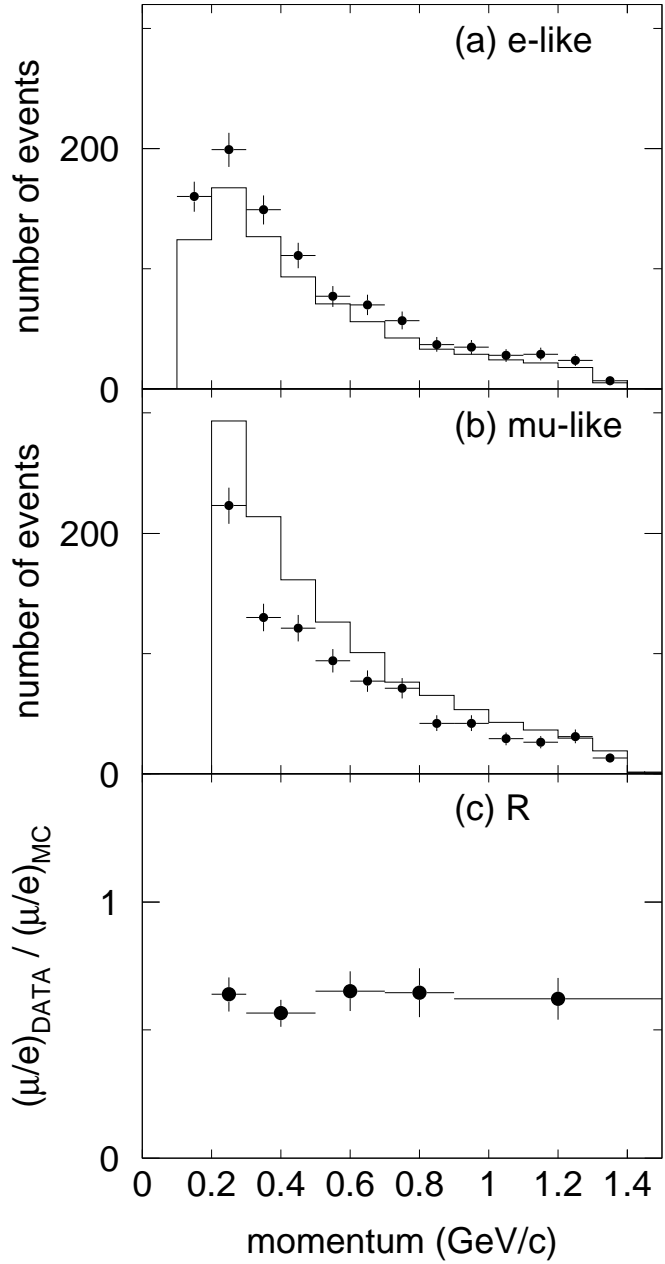}
   \end{center}
   \caption{Momentum distributions for: (a) $e$-like events, (b)
     $\mu$-like events, and (c) $R$.  The histograms show the Monte
     Carlo prediction. Error bars represent statistical errors only.}
    \label{fig:mom}
\end{figure}

Figures~\ref{fig:mom}(a) and (b) show the momentum distributions of
the $e$-like and $\mu$-like events, respectively.
The systematic uncertainty in the
absolute normalization of the Monte Carlo events is $\pm$25\%: 20\%
from the uncertainty of the neutrino flux calculation and 15\% from
the neutrino interaction cross section. 
As a result, we cannot determine from 
these data alone whether the observed deviation of $R$ from unity is
due to an electron excess or a muon deficit.
The shape of each
distribution was consistent with Monte Carlo prediction; $\chi^2/{\rm
d.o.f.}$ was $2.9/12$ for the $e$-like events and $12.2/12$ for
$\mu$-like events.  Figure~\ref{fig:mom}(c) shows $R$ as a function of
momentum.  It is consistent with a flat distribution within the
statistical error.

\begin{figure}
   \begin{center}
      \epsfxsize=10cm
      \epsfbox{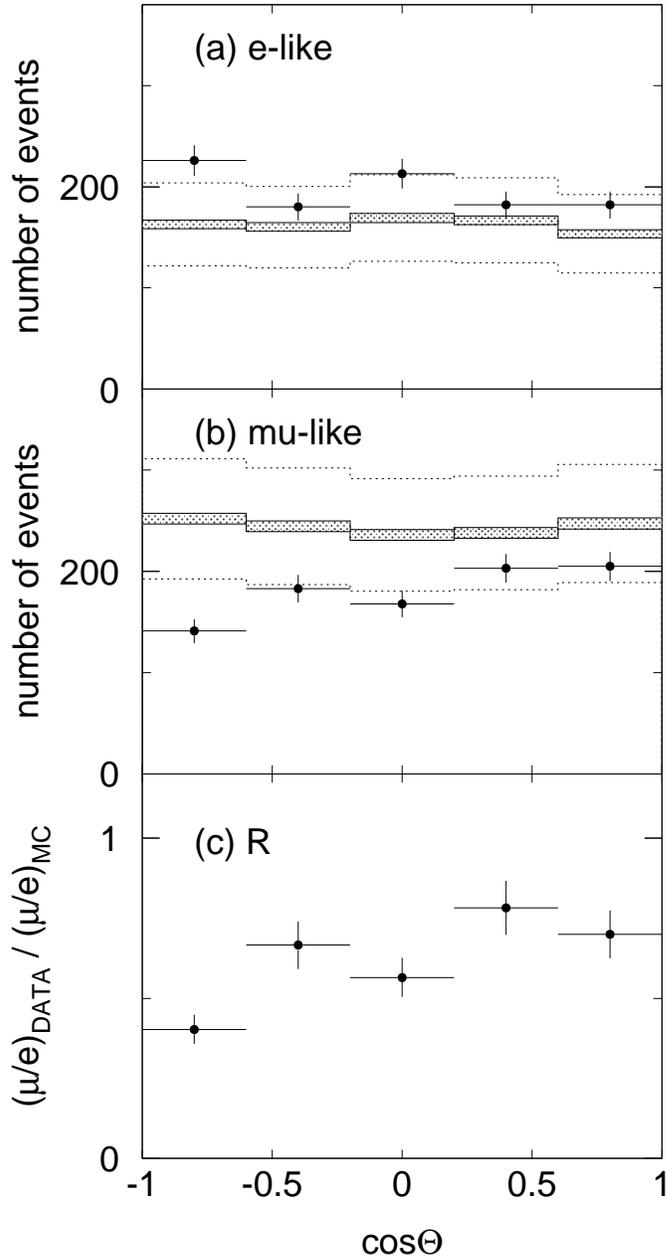}
   \end{center}
   \caption{Zenith angle distributions for: (a) $e$-like events,
             (b) $\mu$-like events, and (c) $R$.
             ($\cos\Theta=1$ means down-going.)
             Histograms with shaded error bars show 
             the Monte Carlo prediction with its statistical error.
             Dotted histograms show the $\pm$25\% systematic uncertainty on the
             absolute normalization, which is correlated between
             $\mu$-like and $e$-like events.}
   \label{fig:zen}
\end{figure}

For the sub-GeV single-ring sample, Monte Carlo studies showed the
mean neutrino energy for CC interactions to be about 700 MeV for
$e$-like events and 800 MeV for $\mu$-like events; the mean angular
correlation between the charged lepton and the neutrino was estimated
to be $54^\circ$ for muons and $62^\circ$ for electrons.
Figures~\ref{fig:zen}(a) and~(b) show the $\cos\Theta$ distributions
for $e$-like and $\mu$-like events, where $\Theta$ is the zenith angle
of the particle direction.  The $\pm$25\% error of normalization is
also shown.  The shape of the distribution was consistent with
expectation for the $e$-like events ($\chi^2/{\rm d.o.f.}=6.5/4$).
However, it was worse ($\chi^2/{\rm d.o.f.}=18.6/4$) for the the
$\mu$-like events.  Figure~\ref{fig:zen}(c) shows $R$ binned by zenith
angle.  Using the two calculated fluxes\cite{ref:honda,ref:gaisser}
and comparing the ($e$-like)$_{MC}$, ($\mu$-like)$_{MC}$, and
($\mu/e$)$_{MC}$ shapes for the five $\cos\Theta$ bins, we found that
the two calculations had $\pm$2.2\% ($\pm$1.4\%) difference for the
$e$-like ($\mu$-like) prediction.  However, they had very similar
$(\mu/e)_{MC}$ vs $\cos\Theta$ distributions.  We conclude that any
up-down systematic asymmetry in $R$ from the uncertainty in the
assumed flux model is less than $\pm$1\%.  We estimated that the
measured energy was $3$\% higher for down-going compared to up-going
particles by studying decay electrons from stopping cosmic ray muons.
This gain asymmetry caused $\pm$0.1\% ($\pm$0.4\%) up-down asymmetry
in $e$-like ($\mu$-like) events, implying an up-down asymmetry in $R$
of $\pm$0.4\%.  The contamination of non-neutrino background, less
than 0.5\%, could have directional correlation and could cause a
maximum of $\pm 1$\% up-down systematic error. Adding these in
quadrature, the systematic uncertainty in the up-down asymmetry in $R$
is 1.5\%.  This systematic uncertainty is negligibly small compared
with the statistical errors in Figure~\ref{fig:zen}(c).

    \section*{Conclusions}

The first measurements of atmospheric neutrinos in the
Super-Kamiokande experiment have confirmed the existence of a smaller
atmospheric $\nu_\mu/\nu_e$ ratio than predicted.  We
obtained $R = 0.61 \pm 0.03(stat.) \pm 0.05(sys.)$ for events in
the sub-GeV range.  The Super-Kamiokande detector has much greater
fiducial mass and sensitivity than prior experiments.  Given the
relative certainty in this result, statistical fluctuations can no
longer explain the deviation of $R$ from unity. 

We gratefully acknowledge the cooperation of the Kamioka Mining and
Smelting Company. The Super-Kamiokande experiment was built and
operated from funding by the Japanese Ministry of Education, Science,
Sports and Culture, and the United States Department of Energy.

    %
%
%

    %
%
%
%
%
%
%
%

    \onecolumn
\end{document}